\documentclass[prd,amsmath,nofootinbib,amssymb]{revtex4}
\usepackage{graphicx}
\usepackage{subfig}
\usepackage{amssymb}
\usepackage{hyperref}
\usepackage[utf8]{inputenc}
\usepackage[english]{babel}
\usepackage{epsfig}
\usepackage{wasysym}
\usepackage{color,xcolor}
\usepackage{pdflscape}
\usepackage{amsmath}
\usepackage{bm}
\usepackage{epsfig}
\usepackage{amsfonts}
\usepackage{dcolumn}
\usepackage{appendix}

\begin{document}

\title{A regular metric does not ensure the regularity of spacetime\\}

\author{Manuel E. Rodrigues$^{(1,2)}$\footnote{E-mail 
address: esialg@gmail.com}, Henrique A. Vieira$^{(1)}$\footnote{E-mail 
address: henriquefisica2017@gmail.com}
}

\affiliation{$^{(1)}$Faculdade de F\'{i}sica, Programa de P\'{o}s-Gradua\c{c}\~{a}o em F\'{i}sica, Universidade Federal do Par\'{a}, 66075-110, Bel\'{e}m, Par\'{a}, Brazil\\
$^{(2)}$Faculdade de Ci\^{e}ncias Exatas e Tecnologia, Universidade Federal do Par\'{a} Campus Universit\'{a}rio de Abaetetuba, 68440-000, Abaetetuba, Par\'{a}, 
Brazil\\
}

\begin{abstract}
In this paper we try to clarify that a regular metric can generate a singular spacetime. Our work focuses on a static and spherically symmetric spacetime in which regularity exists when all components of the Riemann tensor are finite. There is work in the literature that assumes that the regularity of the metric is a sufficient condition to guarantee it. We study three regular metrics and show that they have singular spacetime.
We also show that these metrics can be interpreted as solutions for black holes whose matter source is described by nonlinear electrodynamics.
We analyze the geodesic equations and the Kretschmann scalar to verify the existence of the curvature singularity. Moreover, we use a change of the line element $r \rightarrow \sqrt{r^2+a^2}$, which is a process of regularization of spacetime already known in the literature. We then recompute the geodesic equations and the Kretschmann scalar and show that all metrics now have regular spacetime. This process transforms them into black-bounce solutions, two of which are new.  We have discussed the properties of the event horizon and the energy conditions for all models.

\end{abstract}

\date{\today}

\maketitle


\section{Introduction}
\label{sec1}
Since the advent of general relativity \cite{Einstein2,Einstein1905}, black holes have always fascinated the scientific community. Classically, these celestial bodies are defined as a region of spacetime covered by an event horizon \cite{Penrose:1969pc} from which not even light can escape. Already in the first solution for a black hole, found by K. Schwarzschild \cite{SZ}, the problem of singularity appears. Singularities are regions of spacetime where physical quantities, such as the density of matter or the curvature, become infinite.
There is heated debate among physicists about whether or not singularities exist in nature, but today there is widespread agreement that this is a possible error in classical theory and that a quantum theory of gravity would not have such an error. Although this area has remained very much in the theoretical realm over the last century, the technological advances of the 21st century have brought hope and euphoria with the discovery of gravitational waves\cite{LIGO1,LIGO2,LIGO3,LIGO4,LIGO5,LIGO6,LIGO7}, and with the publication of the first pictures of the shadow of a black hole by the Event Horizon Telescope Team \cite{fotoBN1,fotoBN2,fotoBN3,fotoBN4,fotoBN5,fotoBN6}.

General relativity and similar theories use Riemannian geometry. In this geometry  the definition of the singularity of spacetime may vary depending on the number of dimensions, symmetries considered, and other aspects. For a static and spherically symmetric spacetime, which may or may not be asymptotically flat, a curvature singularity is when one or more components of the Riemann tensor are infinite \cite{Rubin,Inverno}. A solution that has no singularities is called regular. The solution proposed by James Bardeen \cite{Bardeen1} has recently attracted increasing attention because it is regular.
With the contribution of Beato and Garcia \cite{Beato}, Bardeen's model is now a static and spherically symmetric exact solution of the Einstein equations, where the source is a self-gravitating magnetic charge described by nonlinear electrodynamics \cite{Born}. It was later realized that nonlinear electrodynamics can be used to construct regular black holes \cite{regular4}, and today there are a variety of regular Bardeen-type solutions. Some that have electric charge as a source \cite{regular2,regular5} and others that also have angular momentum \cite{regular6,regular7}.
 Moreover, the solution was extended to alternative theories of gravity, such as the
$f(r)$-theory \cite{regular1,regular3,regular9}, the $f(G)$-theory \cite{regular9,regular8}, and rainbow gravity \cite{regular10}. There are also papers in the literature showing how the regularity of the model is lost when a singular solution is attached to the model \cite{Henrique,Rodrigues:2022zph}.
Besides the ones mentioned here, there are many other regular solutions. In Cosmology, there are sudden singularities \cite{cosmology1,cosmology2,cosmology3}, where in the FLRW metric, the scale factor $a(t)$, remains finite for a finite time $t_s$, where the curvature scalars diverge. This is a well-known example where the metric is regular and the spacetime has a singularity.

On general relativity, there are some papers \cite{Balart:2014cga,Zhang:2022tpr,Zaslavskii:2010qz,Bronnikov:2006fu,Ansoldi:2008jw} which use a metric regularity condition which is necessary but not sufficient to guarantee that all coefficients of the Riemann tensor are finite.
 Their argument is as follows. Given the general metric of a static and spherically symmetric spacetime
\begin{equation}
    ds^2 = f(r) dt^2 - \frac{dr^2}{f(r)} - \Sigma^2(r) \left( d \theta^2 + \sin^2 \theta d \phi^2   \right),
    \label{eq:ds}
\end{equation}
with
\begin{equation}
    f(r) = 1-\frac{M(r)}{r},
    \label{fr}
\end{equation}
and
\begin{equation}
    \Sigma (r) = r.
\end{equation}

They pointed out that the condition of regularity of space-time is \cite{Balart:2014cga,Bronnikov:2006fu,Ansoldi:2008jw}
\begin{equation}
    \lim_{r \to 0} \frac{M(r)}{r}  = k_1,
    \label{eq:con1}
\end{equation}
and
\begin{equation}
 \lim_{r \to \infty} \frac{1}{2} M (r) =  \lim_{r \to \infty} M_{HMS}(r) = k_2,
 \label{eq:con2}
\end{equation}
where $k_1$ and $k_2$ must be a finite value independent of $r$ or zero and $ M_{HMS}(r)$ is the Hernandez–Misner–Sharp quasi-local mass defined by \cite{Lobo:2020ffi}
\begin{equation}
    M_{HMS}(r) = \frac{1}{2} \Sigma (r) \biggl[ 1 - f(r) \left( \frac{d \Sigma (r)}{dr} \right)^2  \biggr].
    \label{MHMS}
\end{equation}

Any function $f(r)$ satisfying the conditions \eqref{eq:con1} and \eqref{eq:con2} has no singularity for any value of $r$.
However, are these conditions sufficient to guarantee that all components of the Riemann tensor are finite? To answer this question, we write this tensor on a general metric given by \eqref{eq:ds}, where $f(r)$ and $\Sigma (r)$ are initially an arbitrary function of the variable $r$. Definition of the Christoffel symbols
\begin{equation}
    \Gamma^{\sigma}_{ \mu \nu} = \frac{1}{2} g^{\sigma \alpha} \left(  \partial_{\nu} g_{\alpha \mu}    + \partial_{\mu} g_{\alpha \nu} - \partial_{\alpha} g_{\mu \nu}          \right),
\end{equation}
and, the most important tensor in the context of general relativity, the Riemann tensor is
\begin{equation}
    R^{\sigma}_{\ \mu \nu \rho} = \partial_{\nu} \Gamma^{\sigma}_{ \mu \rho} - \partial_{\rho} \Gamma^{\sigma}_{ \mu \nu} +\Gamma^{\beta}_{ \mu \rho} \Gamma^{\sigma}_{ \beta \nu} - 
    \Gamma^{\beta}_{ \mu \nu} \Gamma^{\sigma}_{ \beta \rho}.
\end{equation}
It is this tensor which indicates the presence of curvatures. For a flat spacetime all components of the Riemann tensor are zero \cite{Rubin}, and as we said before, for a curvature singularity free spacetime all components are finite. The line element \eqref{eq:ds} leads to the non-zero components of the Riemann tensor
\begin{equation}
    R^{01}_{\ \ 01} = \frac{1}{2} f^{\prime \prime}, \ \ \  R^{02}_{\ \ 02} =  R^{03}_{\ \ 03} = \frac{f^{\prime} \Sigma^{\prime}}{2 \Sigma}, \ \ \  R^{12}_{\ \ 12} =  R^{13}_{\ \ 13} =  \frac{f^{\prime} \Sigma^{\prime} + 2 f \Sigma^{\prime \prime}}{2 \Sigma}, \ \ \    R^{23}_{\ \ 23} = \frac{f (\Sigma^{\prime})^2 -1}{\Sigma^2}.
    \label{Rieaman}
\end{equation}

Where the prime denotes derivation with respect to radial coordinate. From this tensor we can construct several invariants, such as  $R_{\mu \nu}R^{\mu \nu}$, $R_{\mu \nu \alpha \beta}R^{\mu \alpha}R^{\nu \beta}$ and others. For regular space-time, all curvature invariants are finite everywhere. rewriting \eqref{Rieaman} in terms of \eqref{fr} we have
\begin{equation}
    R^{01}_{\ \ 01} = \frac{1}{2} \left( -\frac{M''}{r}+\frac{2 M'}{r^2}-\frac{2 M}{r^3} \right),
\end{equation}
\begin{equation}
     R^{02}_{\ \ 02} =  R^{03}_{\ \ 03} = \frac{1}{2} \left( \frac{M \Sigma '}{r^2 \Sigma }-\frac{M' \Sigma '}{r \Sigma }  \right),
\end{equation}
\begin{equation}
    R^{12}_{\ \ 12} =  R^{13}_{\ \ 13} =  \frac{1}{2} \left( \frac{M \Sigma '}{r^2 \Sigma }-\frac{M' \Sigma '}{r \Sigma }  \right) + \frac{\Sigma ''}{\Sigma }-\frac{M \Sigma ''}{r \Sigma },
\end{equation}
\begin{equation}
     R^{23}_{\ \ 23} =\frac{\Sigma '^2}{\Sigma ^2}-\frac{M \Sigma '^2}{r \Sigma^2} - \frac{1}{\Sigma^2}.
\end{equation}
As one can easily verify from the above equations, the conditions \eqref{eq:con1} and \eqref{eq:con2} are not sufficient to prevent curvature singularities.
We believe that this kind of reasoning in some articles may lead readers, especially inexperienced readers, to misunderstand the singularity of spacetime.

The most concise way to verify this is to calculate the Kretschmann scalar $R_{\mu \nu \alpha \beta}R^{\mu \nu \alpha \beta}$, which can be written as the sum of the squares of all non-zero components of the Riemann tensor \cite{Rubin}
\begin{equation}
    K = 4 (R^{01}_{\ \ 01})^2 + 8 (R^{02}_{\ \ 02})^2+ 8 (R^{12}_{\ \ 12})^2 + 4 (R^{23}_{\ \ 23})^2.
    \label{eq:Riemankre}
\end{equation}
In terms of the functions $f(r)$ and $\Sigma(r)$ we have
\begin{equation}
K =  \frac{\left( \Sigma^2 f^{\prime \prime} \right)^2 + 2 \left(\Sigma f^{\prime} \Sigma^{\prime} \right)^2 +2 \Sigma^2 \left( f^{\prime} \Sigma^{\prime} + 2 f \Sigma^{\prime \prime} \right)^2 +4 \left( 1 - f (\Sigma^{\prime})^2 \right)^2}{\Sigma^4}.
\end{equation}
 A sum of finite quantities is by no means an infinite quantity. Therefore, we consider the equation \eqref{eq:Riemankre} and conclude \cite{Lobo:2020ffi} that the finiteness of the scalar ensures the regularity of spacetime. In the appendix \ref{sec:invariantes} we have commented on the possibility of using the invariants $R_{\mu \nu}R^{\mu \nu}$ and $R_{\mu \nu}g^{\mu \nu}$.

The goal of this paper is to illustrate that a regular metric can lead to a singular spacetime. We will deal only with the case of general relativity, without considering alternative theories of gravity.
We will use three metrics with this property as examples. We will show that they are exact solutions of Einstein's equations when we consider a source of matter described by nonlinear electrodynamics. We will also calculate the energy conditions.
Moreover, we will show that the analysis of the Kretschmann scalar is the most concise method to confirm the regularity of spacetime. To demonstrate this, we will compute the \eqref{eq:con1} and \eqref{eq:con2} conditions, the geodesic equations, and the Kretschmann scalar for each example. Additionally, we will regularize the metrics under consideration and recompute the equations of the geodesics and the Kretschmann scalar. These now regularized metrics are black-bounce solutions, two of which have never been published before. We have also established that the source needed to describe them is a scalar field in addition to nonlinear electrodynamics.
The paper is organized as follows: in section \ref{sec:geo} we give a brief summary on geodesic equations; in section \ref{sec3} we show three examples of regular metrics generating a singular spacetime; in section \ref{sec4} we briefly describe a procedure for regularizing spacetime and apply it to the examples in the previous section; in section \ref{sec5} we draw our conclusions. Appendix \ref{sec:source} contains the method for obtaining a source for a static and spherically symmetric metric. In the appendix \ref{sec:invariantes} we compute two more invariants of general relativity (the Ricci scalar and Ricci squared) for the line element under consideration. The Appendix \ref{sec:energy} contains the definitions of the energy conditions that we use in this work.
We will use the metric signature $(,-,-,-)$ in this work. Also, we will use geometrized units where $G = c =1$.

 \section{Geodesic equations}
 \label{sec:geo}
The equations of motion for free particles in general relativity are the so-called geodesic equations. Since they are the generalization of the straight line to curved spacetime, it is expected that they can be extended to all points, so that the sudden end of a geodesic may indicate a singularity. The general form of the equation is \cite{Inverno}

\begin{equation}
    \frac{d^2 x^{\mu}}{ds^2} + \Gamma^{\mu}_{\nu \alpha} \frac{d x^{\nu}}{ds} \frac{dx^{\alpha}}{ds} = 0,
    \label{eq:geo1}
\end{equation}

the parameter $s$ is a privileged one and is called affine parameter.
For a static and spherically symmetric spacetime given by \eqref{eq:ds} and defining the canonically conjugate momenta at coordinates $t$ and $\phi$, respectively, by

\begin{equation}
    E = \frac{d \mathcal{L}}{ d \Dot{t}} = g_{00} \Dot{t},
\end{equation}

\begin{equation}
    L = \frac{d \mathcal{L}}{ d \Dot{\phi}} =  g_{33} \Dot{\phi},
\end{equation}

where $\mathcal{L}$ is the Lagrangian and the dot represents $d/ds$.  Equation \eqref{eq:geo1}  become 

\begin{equation}
    \Dot{r}^2 g^{11}g_{22} \Dot{\theta}^2 - g^{11} \epsilon + \biggl[ \frac{g^{11}}{g_{00}g_{33} }  \biggr] \left( g_{33}E^2+g_{00}L^2  \right) =0.
    \label{eq:geo2}
\end{equation}
Note that the difference between timelike, spacelike and null geodesics are given by
\begin{equation} \label{eq1}
\begin{split}
\epsilon & =1 \ \ \ \ \ \text{for timelike,}  \\
 \epsilon & = 0 \ \ \ \ \ \text{for null,} \\
 \epsilon & = -1 \ \ \ \ \ \text{for spacelike.}
\end{split}
\end{equation}

Considering that the geodesic is located in equatorial plane $\theta= \pi /2$  we have
\begin{equation}
    \frac{g^{11} \left(E^2 g_{33} + g_{00}
   L^2\right)}{g_{00} g_{33}}-g^{11} \epsilon +\Dot{r}^2 = 0.
   \label{eq:geo3}
\end{equation}
Integration of the above equation should result in a real function $r(s) \in \ \mathcal{R} $, so that $s$ can take any real value. If there is a region of spacetime where real values of $s$ generate a complex function, we say that the geodesic in this region is not extensible. For the static and spherically symmetric spacetime we consider, this geodesic incompleteness indicates the presence of a curvature singularity. 

\section{Examples}
\label{sec3}
In this section we will highlight three examples of metric functions which satisfy \eqref{eq:con1} and \eqref{eq:con2}, i.e. are regular everywhere, but they have an infinite Kretschmann scalar at some point and therefore the spacetime is singular. We will consider $\Sigma (r) = r$ for all examples. 
The reason we consider positive and negative values of $r$-coordinate in the figures of this section is that in the next section we will apply a process of regularization of spacetime which makes the three models proposed here black-bounces solutions.

\subsection{Model 1}
\label{subsec1}

\begin{figure}[htpb]
    \centering
    \includegraphics[scale=0.5]{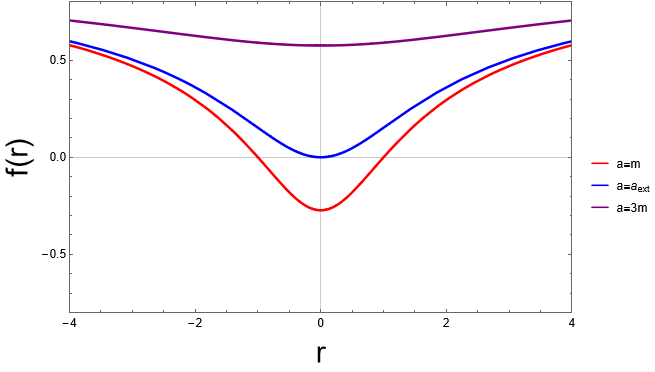}
    \caption{Graphical representation of $f(r)$ with respect to the radial coordinate with the value $m=1$.}
    \label{fig:um}
\end{figure}

\begin{figure}[htpb]
    \centering
    \includegraphics[scale=0.5]{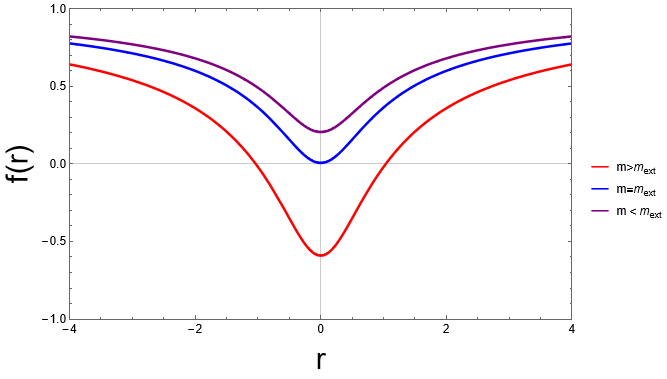}
    \caption{Graphical representation of $f(r)$ with respect to the radial coordinate with the value $a=0.64$.}
    \label{fig:dois}
\end{figure}

For the first example, we will use the metric presented in \cite{Lobo:2020ffi} with $n=1$, which reads

\begin{equation}
    M(r) = \frac{4 m \tan ^{-1}\left(\frac{r}{a}\right)}{\pi },
    \label{eq:massa1}
\end{equation}
where  $m$ is the ADM mass and $a$ is a positive constant, and so
\begin{equation}
    f(r) = 1-\frac{4 m \tan ^{-1}\left(\frac{r}{a}\right)}{\pi  r}.
    \label{eq:metrica1}
\end{equation}
 Although the focus of our work is not to show that the metrics used are solutions of Einstein's equation, it is worth noting that \eqref{eq:metrica1} can be obtained from the field equations of general relativity minimally coupled with nonlinear electrodynamics. The source Lagrangian is (for more details on the determination of the source, see the appendix \ref{sec:source})

\begin{equation}
    L (F) = \frac{F m}{2 \pi ^2 a^2 \sqrt{F}+2 \pi ^2 a}.
    \label{L1}
\end{equation}
Here $F$ is the Maxwell scalar $F = F^{\mu \nu}F_{\mu \nu}/4$  and we identify $a=q$ (where $q$ is a self-gravitating magnetic charge described by nonlinear electrodynamics as mentioned in the appendix).
To check the regularity of the above function we will use the conditions \eqref{eq:con1} and \eqref{eq:con2}. The first one is
\begin{equation}
    \lim_{r \to 0} \frac{M(r)}{r}  = \frac{4 m}{\pi  a},
    \label{eq:rto01}
\end{equation}
which is constant, and  the second one is
\begin{equation}
    \lim_{r \to \infty} \frac{1}{2} M(r) = a m.
\end{equation}
The Hernandez–Misner–Sharp quasi-local mass \eqref{MHMS} for this model is
\begin{equation}
    M_{HMS}(r) = \frac{2 m \tan ^{-1}\left(\frac{r}{a}\right)}{\pi },
\end{equation}
and  its limit when $r \rightarrow \infty$ is $am$. These results prove that the conditions \eqref{eq:con1} and \eqref{eq:con2} are satisfied. Alternatively, we can  expand the metric function in Taylor series to $r<<1$ and $r>>1$  around the value zero
\begin{equation}
    f(r \sim 0)  = 1-\frac{4 m}{\pi  a}+\frac{4 m r^2}{3 \pi  a^3}-\frac{4 m r^4}{5 \left(\pi 
   a^5\right)}+ \mathcal{O}\left(r^5\right),
   \label{eq:24}
\end{equation}
\begin{equation}
     f(r \sim \infty)  = 1-\frac{2 m}{r}+\frac{4 a m}{\pi  r^2}-\frac{4 \left(a^3 m\right)}{3
   \pi  r^4}+ \mathcal{O}\left(\left(\frac{1}{r}\right)^5\right).
   \label{eq:25}
\end{equation}
 Note that for $r>>1$ the variable is $1/r <<1$. From the above two equations we see the asymptotic behavior of the function $f(r)$ and find that it is regular in these limits. In figures \ref{fig:um} and \ref{fig:dois} we show a graphical representation of this metric function.
We note that the $r$-coordinate does not take negative values. The figures are so because in section \ref{sec4} we will convert these models into black-bounces solutions and in this context the negative region is given a physical interpretation. 
The extreme values of the parameters $a$ and $m$ mentioned in the figures can be determined with  the help of the event horizon condition
\begin{equation}
    f(r_+) = 0,
    \label{eq:rh1}
\end{equation}
where $r_+$ represents the radius of the horizon. And furthermore we have to assume the following
\begin{equation}
    \frac{df(r_+)}{dr_+} = 0.
\end{equation}
From this we can derive both the value of the radius of the event horizon and the extreme values of the parameter $a_{ext}$ or $m_{ext}$. Although we could not find an analytical result for $r_+$, we know that $a_{ext} = 4m/\pi$ \cite{Lobo:2020ffi}, and that the solution has only a positive event horizon. Later, we will regularize these examples with Simpson-Visser spacetime and then recover the black-bounce behavior of this metric described in the original source mentioned above.

For completeness, we will also review the energy conditions; in the Appendix \ref{sec:energy} we show the definitions we use to calculate these conditions. Outside the event horizon, i.e., when t is a timelike coordinate, we have

\begin{equation}
    NEC_1 \rightarrow 0 \geq 0, \ \ \ \ \ NEC_2 \rightarrow \frac{a m \left(a^2+2 r^2\right)}{2 \pi ^2 r^2 \left(a^2+r^2\right)^2} \geq 0
\end{equation}
\begin{equation}
    DEC_3 \rightarrow \frac{a m}{2 \pi ^2 r^2 \left(a^2+r^2\right)} \geq 0  \ \ \ \ \ SEC_3 \rightarrow \frac{a m}{\pi ^2 \left(a^2+r^2\right)^2} \geq 0,
\end{equation}
\begin{equation}
    DEC_1 \rightarrow \frac{a m}{\pi ^2 r^2 \left(a^2+r^2\right)} \geq 0   \ \ \ \ \ DEC_2 \rightarrow \frac{a^3 m}{2 \pi ^2 r^2 \left(a^2+r^2\right)^2} \geq 0.
\end{equation}
Since $NEC_1$ is zero, it is easy to see that this condition will always be satisfied (in fact, due to the dependence on the function $\Sigma(r)$, the condition $NEC_1$ is zero for all other models as long as we consider $\Sigma(r) =r$). Note that all other conditions have only an even exponent dependence on the radial coordinate, which means that they are never negative, i.e., they are always satisfied. The same is true for the region inside the event horizon.


Now we will check the completeness of the geodesic equations. We find the geodesic equation for this model using \eqref{eq:metrica1} in \eqref{eq:geo3}, which leads to the following result
\begin{equation}
\Dot{r}^2 +  \frac{4 L^2 m \tan ^{-1}\left(\frac{r}{a}\right)}{\pi  r^3}-\epsilon  \left(\frac{4 m \tan ^{-1}\left(\frac{r}{a}\right)}{\pi
    r}-1\right)-E^2-\frac{L^2}{r^2} =0.
\end{equation}
Our task now is to integrate the above equation to find a general function $r=r(s)$, but this is not an easy task. Therefore, for this and the other examples, we will use the Taylor series approximation for 3 important limits: $r > > 1$, $r < < 1$, $r \rightarrow r_+$. In other words, we will focus on the behavior of $r(s)$ for a distance far from the black hole, at the origin of the coordinate system and at the event horizon. For the first case we have

\begin{equation}
     \Dot{r}^2-E^2+\epsilon +\frac{2  m
   \epsilon }{r}+ \mathcal{O}\left(\left(\frac{1}{r}\right)^2\right) = 0,
   \label{eq:geo4a}
\end{equation}
The term $2m \epsilon /r$ can be regarded as zero, and then by integration 

\begin{equation}
    r(s) \approx m \pm s \sqrt{E^2-\epsilon }+c_1,
\end{equation}

where $c_1$ is a positive constant of integration. From this result we see that the function
$r(s)$ can take any real value of the affine parameter and $s \rightarrow \infty$ leads to $r(s) \rightarrow \infty$, so it is extensible to future infinity. Later we will analyze the Kretschmann scalar for these limits and show that the results are the same.
For $r< < 1$, the expansion of the equation \eqref{eq:geo3} is

\begin{equation}
    \frac{L^2 \left(\frac{4 m}{\pi  a}-1\right)}{r^2}+\frac{-3 \pi 
   a^3 E^2-4 L^2 m}{3 \pi  a^3}+\epsilon  \left(1-\frac{4 m}{\pi
    a}\right)+\Dot{r}^2+ \mathcal{O} \left(r^2\right) = 0,
    \label{eq:geo5a}
\end{equation}
which the integration leads to

\begin{equation}
    r(s) \approx \pm \frac{\sqrt{-L_1-\epsilon_1^2
   (s \pm c_2){}^2}}{\sqrt{\epsilon_1}}.
   \label{rzero1}
\end{equation}
Here $L_1$ and $\epsilon_1$ are

\begin{equation}
    L_1 = L^2 \left(\frac{4 m}{\pi  a}-1\right),
\end{equation}
\begin{equation}
    \epsilon_1 = \epsilon  \left(1-\frac{4 m}{\pi  a}\right).
\end{equation}
And $c_2$ is another constant of integration. From equation \eqref{rzero1} we conclude that $s$ is not extensible to a real value,because if $s \rightarrow \infty$, the function $r(s)$ for $\epsilon_1 > 0$ will not be analytic.So we say that spacetime is geodesically incomplete near the origin. This indicates a singularity and is later reinforced by the analysis of the Kretschmann scalar.

For $r=r_+$ we have

\begin{equation}
    A_1+A_2 \left(  r - r_+   \right) + \Dot{r}^2 + \mathcal{O}\left(\left(r-r_+\right)^2\right) = 0,
    \label{geo4}
\end{equation}
where $A_1$ and $A_2$ are

\begin{equation}
    A_1 = \frac{4 L^2 m \tan ^{-1}\left(\frac{r_+}{a}\right)+\pi 
   \left(-E^2\right) r_+^3-\pi  L^2 r_+}{\pi 
   r_+^3},
\end{equation}
\begin{equation}
    A_2 = \frac{2 \left(-6 a^2 L^2 m \tan
   ^{-1}\left(\frac{r_+}{a}\right)+\pi  a^2 L^2 r_+-6 L^2 m
   r_+^2 \tan ^{-1}\left(\frac{r_+}{a}\right)+2 a L^2 m
   r_++\pi  L^2 r_+^3\right)}{\pi  r_+^4
   \left(a^2+r_+^2\right)}-\frac{4 m \epsilon 
   \left(\frac{a}{r_+ \left(a^2+r_+^2\right)}-\frac{\tan
   ^{-1}\left(\frac{r_+}{a}\right)}{r_+^2}\right)}{\pi }.
\end{equation}

Solving the equation \eqref{geo4}, we find

\begin{equation}
    r(s) \approx \biggl[ - \frac{ A_1}{A_2}   +r_+ - \frac{A_2}{4} \left( s \pm c_3  \right) \biggr].
\end{equation}
This result shows that $s$ can take any value in this limit and therefore spacetime is extensible up to $r=r_+$.

A more concise way to look for curvature singularities in this kind of spacetime is to analyze the Kretschmann scalar. For the metric \eqref{eq:metrica1} we have
\begin{equation}
  K = \frac{4 r^2 \left(\frac{4 m \tan ^{-1}\left(\frac{r}{a}\right)}{\pi  r^2}-\frac{4 m}{\pi  a r
   \left(\frac{r^2}{a^2}+1\right)}\right)^2+r^4 \left(\frac{8 m}{\pi  a r^2 \left(\frac{r^2}{a^2}+1\right)}+\frac{8 m}{\pi 
   a^3 \left(\frac{r^2}{a^2}+1\right)^2}-\frac{8 m \tan ^{-1}\left(\frac{r}{a}\right)}{\pi  r^3}\right)^2+\frac{64 m^2 \tan
   ^{-1}\left(\frac{r}{a}\right)^2}{\pi ^2 r^2}}{r^4}.
\end{equation}
As pointed out in \ref{sec1}, the finiteness of this invariant at a point guarantees that spacetime is regular there. We also show that for $r< < 1$ the geodesics are not extensible. So we expand the Kretschmann scalar in this region
\begin{equation}
 K (r \sim 0)  = \frac{64 m^2}{\pi ^2 a^2 r^4}-\frac{128 m^2}{3 \left(\pi ^2 a^4\right) r^2}+\frac{1024 m^2}{15 \pi ^2 a^6}-\frac{1024 m^2
   r^2}{7 \left(\pi ^2 a^8\right)}+ \mathcal{O}\left(r^3\right).
    \end{equation}
The first two terms are proportional to $1/r^4$ and $1/r^2$, respectively. These tend toward infinity as $r \rightarrow 0$.

With these analyses, we show that the metric function generating spacetime is regular, i.e., it is smooth, derivable, and finite at all points of the radial coordinate. However, spacetime itself has a curvature singularity at the origin. This example underlines the importance of what we said in the introduction: A regular metric can produce a spacetime with singularities. We will now show two more examples with the same property.

\subsection{Model 2}
\label{subsec2}

As our second example we chose the following mass function
\begin{equation}
    M(r) = 2 m r \tan ^{-1}\left(\frac{a}{\sqrt{a^2+r^2}}\right),
\end{equation}
where again $m$ is the ADM mass and $a$ is a positive constant. Now, the metric is 

\begin{equation}
    f(r) = 1-2 m \tan ^{-1}\left(\frac{a}{\sqrt{a^2+r^2}}\right).
    \label{eq:metrica2}
\end{equation}
 As with the previous model, the  metric \eqref{eq:metrica2} is an exact solution of the Einstein equations if we take the Lagrangian 
\begin{equation}
    L(F) = \frac{m \sqrt{F}}{4 \pi}   \left(\frac{\tan ^{-1}\left(\frac{a}{\sqrt{a
   \left(\frac{1}{\sqrt{F}}+a\right)}}\right)}{a}-\frac{1}{\sqrt{a \left(\frac{1}{\sqrt{F}}+a\right)} \left(2
   \sqrt{F} a+1\right)}\right)
   \label{L2}
\end{equation}
Where the parameter $a$ is interpreted as a self-gravitating magnetic charge described by nonlinear electrodynamics.

\begin{figure}[htpb]
    \centering
    \includegraphics[scale=0.5]{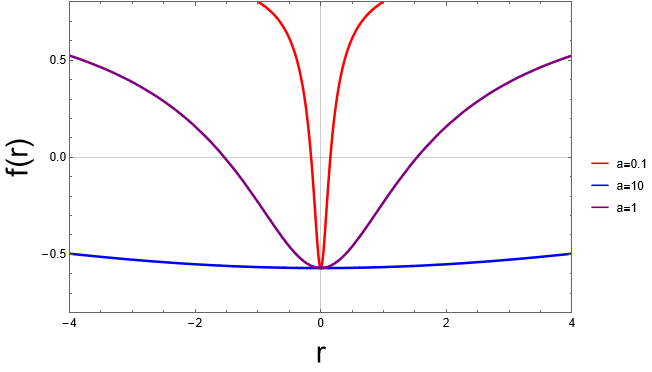}
    \caption{Graphical representation of $f(r)$ with respect to the radial coordinate with the value $m=1$.}
    \label{fig:tres}
\end{figure}

\begin{figure}[htpb]
    \centering
    \includegraphics[scale=0.5]{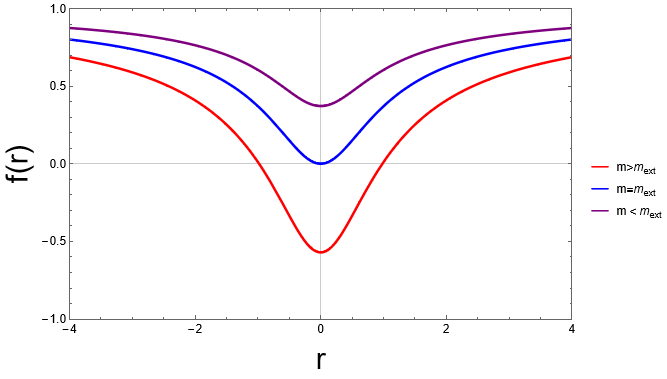}
    \caption{Graphical representation of $f(r)$ with respect to the radial coordinate with the value $a=0.64$.}
    \label{fig:quatro}
\end{figure}

We proceed in the same way as in the first example. First we check the conditions \eqref{eq:con1} and \eqref{eq:con2}. They are
\begin{equation}
    \lim_{r \to 0} \frac{M(r)}{r} = \frac{\pi  m}{2},
\end{equation}
\begin{equation}
      \lim_{r \to \infty} \frac{1}{2}M(r) =  \lim_{r \to \infty} M_{HMS}(r) = a m.
\end{equation}
That is, the regularity conditions of the metric are fulfilled. In addition,
the Taylor series of $f(r)$ function to $r>>1$ and $r<<1$ around the value zero are
\begin{equation}
 f(  r \sim 0)  = 1-\frac{\pi m}{2} +\frac{ m r^2}{2 a^2}-\frac{r^4
   \left( m\right)}{4 a^4}+ \mathcal{O}\left(r^5\right),
\end{equation}
\begin{equation}
    f(  r \sim \infty  ) = 1-\frac{2 a m}{r}+\frac{5 a^3 m}{3 r^3}+\mathcal{O}\left(\left(\frac{1}{r}\right)^5\right).
\end{equation}
These two equations show that the asymptotic behavior of the metric is regular at infinity as well as at the origin. We show the behavior of \eqref{eq:metrica2} in the figures \ref{fig:tres} and \ref{fig:quatro}. This solution has only one positive event horizon, similar to the previous model, but it is not present in \cite{Lobo:2020ffi}. This metric is a new solution.
The event horizon and the extreme value of the mass are given by
\begin{equation}
    r_+ = \cot \left(\frac{1}{2 m}\right) \sqrt{a^2 \cos \left(\frac{1}{m}\right) \sec ^2\left(\frac{1}{2 m}\right)},
\end{equation}
\begin{equation}
    m_{ext} = 0.63662.
\end{equation}

\begin{figure}[htpb]
    \centering
    \includegraphics[scale=0.5]{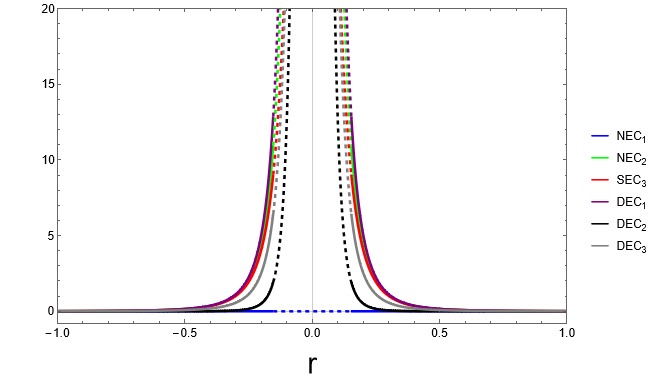}
    \caption{Graphical representation of the energy conditions for the model 2 with $m=1$ and $a=0.01.$ The filled curves represent regions outside and the dashed curves represent regions within the event horizon.}
    \label{fig:energy2}
\end{figure}

Given the more complicated nature of this model, we have chosen to analyze the energy conditions graphically. Figure \ref{fig:energy2} shows the energy conditions inside and outside the event horizon. We can see that all the energy conditions are satisfied.

To calculate the geodesic equation, we again use \eqref{eq:geo3}, which leads us to the following result
\begin{equation}
  \Dot{r}^2  -\frac{-2 L^2 m \tan ^{-1}\left(\frac{a}{\sqrt{a^2+r^2}}\right)+E^2 r^2+L^2}{r^2}-\epsilon  \left(2 m \tan
   ^{-1}\left(\frac{a}{\sqrt{a^2+r^2}}\right)-1\right) = 0.
   \label{eq:geom2}
\end{equation}
Just as in the \ref{subsec1} section, we will not integrate \eqref{eq:geom2} analytically.
We will work with the same limits we used before. For $r> > 1$, the Taylor series is

\begin{equation}
    \Dot{r}^2-E^2+\epsilon -\frac{2 a m \epsilon
   }{r}+\mathcal{O}\left(\left(\frac{1}{r}\right)^2\right) = 0.
\end{equation}
We integrate the above equation and get
\begin{equation}
    r(s) \approx \pm s \sqrt{E^2-\epsilon} +c_4,
\end{equation}
where $c_4$ is simply a positive constant of integration. Since $s \rightarrow \infty$ leads to $r(s) \rightarrow \infty$, we conclude that the function $r(s)$ is extensible to future infinity.
For $r< < 1$ the Taylor series is
\begin{equation}
\frac{L^2 \left(2 m \tan ^{-1}\left(1\right)-1\right)}{r^2}+\left(-\frac{m
   \left(\pi  a^2 \epsilon +L^2\right)}{2 a^2}-E^2+\Dot{r}^2+\epsilon \right)+\mathcal{O}\left(r^2\right) = 0.
   \label{eq:geozero2}
\end{equation}
Note that equation \eqref{eq:geozero2} has the same form of \eqref{eq:geo5a}, therefore the integration lead to the same result
\begin{equation}
    r(s) \approx \pm \frac{\sqrt{-L_2-\epsilon_2^2
   (s \pm c_5){}^2}}{\sqrt{\epsilon_2}}.
   \label{rzero2}
\end{equation}
Here, $c_5$ is a positive integration constant and
\begin{equation}
    L_2 = L^2 \left(2 m \tan ^{-1}\left(1\right)-1\right),
\end{equation}
\begin{equation}
    \epsilon_2 = -\frac{m
   \left(\pi  a^2 \epsilon +L^2\right)}{2 a^2}-E^2+\epsilon.
\end{equation}
Again using the same arguments as model one, we consider the equation \eqref{rzero2} and we can say that the function $r(s)$ is not extensible to every single value of $s$. Finally, for $r = r_+$ we have the Taylor series of the geodesic equation, which is similar for model 1 \eqref{geo4}. Thus, the conclusion is the same. The function $r(s)$ is extensible to $r= r_+$. From these analyzes we conclude that the spacetime generated by the metric \eqref{eq:metrica2} is geodesically incomplete for$r \rightarrow 0$.

In summary, the Kretschmann scalar for metric \eqref{eq:metrica2} is

\begin{equation}
    K = \frac{16 m^2 \tan ^{-1}\left(\frac{a}{\sqrt{a^2+r^2}}\right)^2+\frac{16 a^2 m^2 r^4}{\left(a^2+r^2\right)^3
   \left(\frac{a^2}{a^2+r^2}+1\right)^2}+\frac{4 m^2 r^4 \left(2 a^5-a^3 r^2-2 a r^4\right)^2}{\left(a^2+r^2\right)^3 \left(2
   a^2+r^2\right)^4}}{r^4},
\end{equation}
 whose asymptotic behavior near the origin can be illustrated by a Taylor series expansion for $r<< 1$
\begin{equation}
  K (r \sim 0) = \frac{\pi ^2 m^2}{r^4}-\frac{2 \left(\pi  m^2\right)}{a^2 r^2}+\frac{(6+\pi ) m^2}{a^4}-\frac{\left((180+7 \pi ) m^2\right)
   r^2}{12 a^6}+\mathcal{O}\left(r^3\right).
\end{equation}

We call attention once again to the terms proportional to $1/r^4$ and $1/r^2$. These terms tend to infinity when $r\rightarrow 0$, i.e., there is a curvature singularity at the origin. So this was another example where we used a regular metric satisfying \eqref{eq:con1} and \eqref{eq:con2}, but this was not enough to guarantee the regularity of spacetime.

\subsection{Model 3
\label{subsec3}}
As a last example we chose
\begin{equation}
    M(r) = \frac{2 m r^2}{a^2+r^2},
\end{equation}
where  $a$ is a positive constant. So the metric is
\begin{equation}
    f(r) = 1-\frac{2 m r}{a^2+r^2}.
    \label{eq:metrica3}
\end{equation}

Like the previous models, this one can be understood as an exact solution of the Einstein equations if we use the following nonlinear Lagrangian as a source
\begin{equation}
    L = \frac{a^2 F^{5/4} m}{2 \pi  \sqrt{a} \left(a^2 \sqrt{F}+a\right)^2},
    \label{L3}
\end{equation}
in addition we take $a=q$ as the magnetic charge described in the appendix  \ref{sec:source}.

\begin{figure}[htpb]
    \centering
    \includegraphics[scale=0.5]{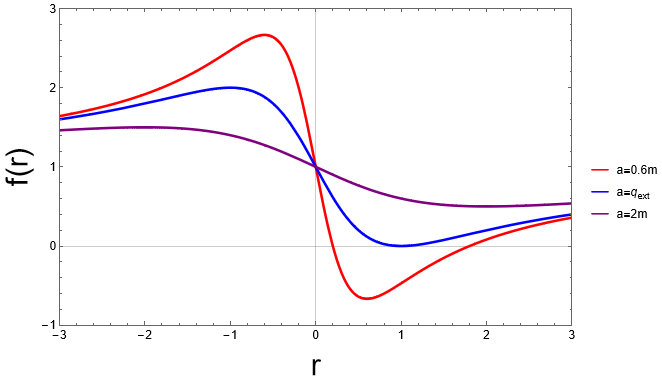}
    \caption{Graphical representation of $f(r)$ with respect to the radial coordinate with the value $m=1$.}
    \label{fig:cinco}
\end{figure}

\begin{figure}[htpb]
    \centering
    \includegraphics[scale=0.5]{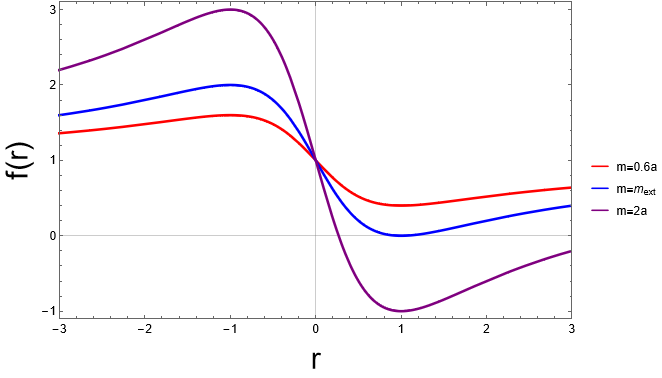}
    \caption{Graphical representation of $f(r)$ with respect to the radial coordinate with the value $a=0.64$.}
    \label{fig:seis}
\end{figure}

The conditions \eqref{eq:con1} and \eqref{eq:con2} are
\begin{equation}
    \lim_{r \to 0} \frac{M(r)}{r} = 0,
\end{equation}
\begin{equation}
      \lim_{r \to \infty} \frac{1}{2}M(r) =  \lim_{r \to \infty} M_{HMS}(r) =  m.
\end{equation}
then the metric regularity conditions mentioned in the literature are fulfilled. Moreover, we can see that the asymptotic behavior of $f(r)$ at the origin and at infinity is the same
\begin{equation}
    f(r \sim 0) = 1-\frac{2 m r}{a^2}+\frac{2 m r^3}{a^4}+\mathcal{O}\left(r^5\right),
\end{equation}
\begin{equation}
    f( r \sim \infty) = 1-\frac{2 m}{r}+\frac{2 m a^2}{r^3}+\mathcal{O}\left(\left(\frac{1}{r}\right)^5\right),
\end{equation}
are in fact regular. We represent the metric function \eqref{eq:metrica3} in the figures \ref{fig:cinco} and \ref{fig:seis}. We see that there are up to two positive event horizons defined by
\begin{equation}
    r_+ = m \pm \sqrt{m^2-a^2},
\end{equation}
where it is easy to see that the extreme solution is reached when
\begin{equation}
    a_{ext} =  m.
\end{equation}

The energy conditions for the region outside the horizon are
\begin{equation}
    NEC_1 \rightarrow 0 \geq 0, \ \ \ \ \ NEC_2 \rightarrow \frac{a^2 m \left(a^2+5 r^2\right)}{4 \pi  r \left(a^2+r^2\right)^3} \geq  0
\end{equation}
\begin{equation}
    DEC_3 \rightarrow \frac{a^2 m}{2 \pi  r \left(a^2+r^2\right)^2}  \geq 0  \ \ \ \ \ SEC_3 \rightarrow  -\frac{m \left(a^4-3 a^2 r^2\right)}{2 \pi  r \left(a^2+r^2\right)^3} \geq 0,
\end{equation}
\begin{equation}
    DEC_1 \rightarrow  \frac{a^2 m}{\pi  r \left(a^2+r^2\right)^2} \geq 0   \ \ \ \ \ DEC_2 \rightarrow \frac{a^2 m \left(3 a^2-r^2\right)}{4 \pi  r \left(a^2+r^2\right)^3}  \geq 0.
\end{equation}
Here we note that $SEC_3$ is violated if $a^2 > 3r^2$, and $DEC_2$ is violated if $3a^2< r^2$. The other conditions are always satisfied. The same is true for the region inside the event horizon (This is because $\Sigma (r)= r$ implies that $\Sigma^{\prime \prime}(r) = 0$, making the energy conditions inside and outside the horizon the same).

As in the previous cases, the geodesic equation we must consider is \eqref{eq:geo3}, which in this case, considering \eqref{eq:metrica3}, becomes
\begin{equation}
\Dot{r}^2-E^2-\frac{L^2 \left(r (r-2 m)+q^2\right)}{r^2 \left(q^2+r^2\right)}-\epsilon  \left(\frac{2 m
   r}{q^2+r^2}-1\right)    =0.
   \label{eq:69}
\end{equation}

This is much simpler than the geodesic equations in the previous cases, but to maintain the same standard of analysis, we will not integrate this equation analytically. We will study the motion of free-falling particles near infinity, the origin, and the event horizon. For the first limit, the evolution of \eqref{eq:69} for $r> > 1$
\begin{equation}
    \Dot{r}^2-E^2+\epsilon -\frac{2 m \epsilon }{r}+\mathcal{O}\left(\left(\frac{1}{r}\right)^2\right) =0,
\end{equation}
and  the integration over all values of the affine parameter leads to
\begin{equation}
    r(s) \approx \pm s \sqrt{E^2-\epsilon} +c_6,
\end{equation}
where $c_6$ is a positive constant of integration. From this result we see that the function $r(s)$ is infinitely extensible in the future for any value of s. For $r< < 1$:
\begin{equation}
\Dot{r}^2-\frac{L^2}{r^2}+\frac{2 L^2 m}{a^2 r}-E^2+\epsilon -\frac{2 r \left(m
   \left(L^2+a^2 \epsilon \right)\right)}{a^4}+\mathcal{O}\left(r^2\right) =0,
\end{equation}
which, when solved, leads to six different solutions, but all of them have the following term
\begin{equation}
\sqrt{\frac{L^2 (s+c_7 a){}^2 \left(3 c_7 L m^2 a+3 L m^2 s-a^4\right) \left(3 c_7 L m^2 a+3 L m^2
   s+a^4\right)}{m^2 a^4}}.
\end{equation}
Here $c_7$ is simply an integration constant. Note that this term can take complex values depending on the variable $s$. Since all solutions we get from $r(s)$ have this term, we see that the affine parameter is not expandable to a real value. In the general case, we now have two event horizons. For $r=r_+ = m + \sqrt{m^2-q^2}$:
\begin{equation}
    \Dot{r}^2-E^2+\left(r-m-\sqrt{m^2-a^2}\right) \left(\frac{ \epsilon  \left(a^2-2
   m^2\right)-\sqrt{m^2-a^2} \left(L^2+2 m^2 \epsilon \right)}{\left(\sqrt{m^2-a^2}+m\right)^3}+ \frac{\epsilon}{m}
   \right)+\mathcal{O}\left(\left(r-r_+\right)^2\right) = 0,
\end{equation}
which has the same structure as the equation \eqref{geo4}. For the other horizon, $r_+ = m - \sqrt{m^2-a^2} $, the same holds. So we know that the function $r(s)$ at the horizons is extensible to any value of the variable $s$. So we can say that spacetime is geodesically complete at infinity and at the event horizon, but not at the origin

 We will now check  the  Kretschmann scalar. For metric \eqref{eq:metrica3} it is
\begin{equation}
    K = \frac{\frac{16 m^2 r^2}{\left(a^2+r^2\right)^2}+4 r^2 \left(\frac{4 m r^2}{\left(a^2+r^2\right)^2}-\frac{2
   m}{a^2+r^2}\right)^2+r^4 \left(\frac{12 m r}{\left(a^2+r^2\right)^2}-\frac{16 m
   r^3}{\left(a^2+r^2\right)^3}\right)^2}{r^4}.
\end{equation}
As with the previous models, we consider the expansion in Taylor series for $r< < 1$
\begin{equation}
 K (r \sim 0 ) =   \frac{32 m^2}{a^4 r^2}-\frac{128 m^2}{a^6}+\frac{496 m^2 r^2}{a^8}+O\left(r^3\right).
\end{equation}

Note that the first term tends to infinity when $r\rightarrow 0$.The Kretschmann scalar diverges at the origin and this means that spacetime is singular, although the metric is regular for all real values of the radial coordinate. So far, we have shown three examples where the conditions \eqref{eq:con1} and \eqref{eq:con2}, found in some works in the literature, were not sufficient to guarantee the finiteness of the components of the Rieman tensor, i.e., the regularity of a static and spherically symmetric spacetime. This was the main goal of our work and we believe we have achieved it with this section. In the next section we show a possible way to regularize the spacetime of the metrics used so far.


\section{Regularization of the singular time-space
\label{sec4}}
In this section  we will apply a spacetime regularization process described in \cite{Simpson:2018tsi,Franzin:2021vnj,Bronnikov:2022bud} to our examples from the last section. It consists of making a change in the equation the equation  \eqref{eq:ds} by doing  $r\rightarrow \sqrt{r^2+a^2}$, in other words, the function $\Sigma(r)$ becomes
\begin{equation}
    \Sigma(r)  = \sqrt{a^2+r^2},
\end{equation}
where $a$ is the same positive constant we use for our metric functions. Note that this also implies a change in the area of the black hole from $\pi r^2$ to $\pi \left( a^2r^2 \right) $.
This change does not affect the regularity of the metric or the event horizon property.
For this reason, we will only repeat the analysis of the geodesic equations and the Kretschmann scalar.  However, this change profoundly affects the energy conditions, since now $\Sigma^{\prime \prime} (r) = \frac{a^2}{\left(a^2+r^2\right)^{3/2}}$. Thus, not only the functions become more extensive, but also the conditions inside and outside the horizon can be different.

\subsection{Model 1}
\begin{figure}[htpb]
    \centering
    \includegraphics[scale=0.5]{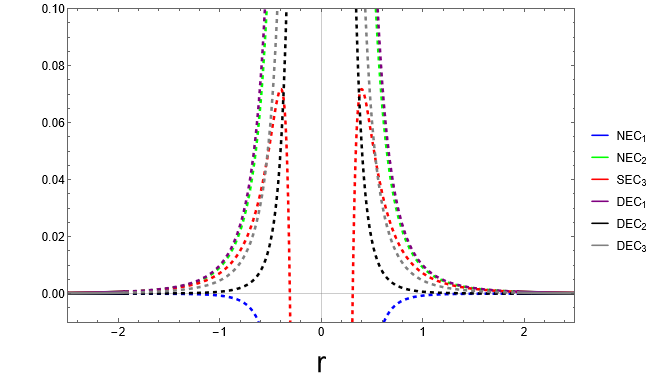}
    \caption{Graphical representation of the energy conditions for regularized model 1 with $m=1$ and $a=0.01$. The filled curves represent the region outside and the dashed curves the region inside the event horizon.}
    \label{fig:energy4}
\end{figure}

The spherically symmetric metric is 
\begin{equation}
    ds^2 = \Biggl[1-\frac{4 m \tan ^{-1}\left(\frac{r}{a}\right)}{\pi  r} \Biggr]    dt^2 - \frac{dr^2}{\Biggl[1-\frac{4 m \tan ^{-1}\left(\frac{r}{a}\right)}{\pi  r} \Biggr]  } - \left(a^2+r^2 \right)  \left( d \theta^2 + \sin^2 \theta d \phi^2   \right).
    \label{eq:ds2}
\end{equation}
We cannot construct a source for the models in Simpson-Visser spacetime as we did before. Now, in addition to the magnetic charge described by NED, we need a scalar field \cite{Rodrigues:2023vtm}. How to find the scalar field and its associated potential given the metric is described in the appendix \ref{sec:source}.
For \eqref{eq:ds2} it holds that

\begin{equation}
    \Phi(r) = \frac{\tan ^{-1}\left(\frac{r}{a}\right)}{2 \sqrt{2 \pi }},
    \label{Phi}
\end{equation}

\begin{equation}
\begin{aligned}
    & V(\Phi) = \frac{m }{8 \pi ^2 a^3 } \Biggl[  \left(\cos \left(4 \sqrt{2 \pi } \Phi \right)+7\right) \cos ^2\left(2 \sqrt{2 \pi } \Phi \right)+\tan
   ^{-1}\left(\tan \left(2 \sqrt{2 \pi } \Phi \right)\right) \Bigl( 6 \tan ^{-1}\left(\tan \left(2 \sqrt{2 \pi
   } \Phi \right)\right) \\ &+9 \cot \left(2 \sqrt{2 \pi } \Phi \right)-\cos \left(6 \sqrt{2 \pi } \Phi \right)
   \csc \left(2 \sqrt{2 \pi } \Phi \right)\Bigr)  \Biggr],
   \end{aligned}
\end{equation}
and

\begin{equation}
    L (F) = \frac{m }{4 \pi ^2 a^3} \left((F-3) F-3 \tan ^{-1}\left(\frac{\sqrt{1-F}}{\sqrt{F}}\right)^2+\frac{2 (F-3) \sqrt{F} \tan
   ^{-1}\left(\frac{\sqrt{1-F}}{\sqrt{F}}\right)}{\sqrt{1-F}}\right).
\end{equation}
Here we use that $\epsilon= -1$, so that the scalar field is real. Moreover, the scalar field depends only on $\Sigma(r)$ and therefore will be the same for the next models. 
The energy conditions inside and outside the event horizon are shown in the figure \ref{fig:energy4}. As we can see, $SEC_3$ and $NEC_1$ can be violated inside the event horizon.

Considering the components of the metric \eqref{eq:ds2} and the equation \eqref{eq:geo3}  we  have
\begin{equation}
    \frac{\pi  r \left(-\left(a^2+r^2\right) \left(E^2-\Dot{r}^2-\epsilon \right)-L^2\right)+4 m \tan
   ^{-1}\left(\frac{r}{a}\right) \left(L^2-\epsilon  \left(a^2+r^2\right)\right)}{\pi  r
   \left(a^2+r^2\right)} = 0,
   \label{eq:geoone}
\end{equation}
this is the geodesic equation for model 1. As in the previous section, we will use the Taylor series expansion to calculate the behavior near infinity, the origin, and the event horizon.

For $r>>1$ the Taylor series of \eqref{eq:geoone} is
\begin{equation}
   \Dot{r}^2 -E^2+\epsilon +\frac{2  m \epsilon
   }{r}+\mathcal{O}\left(\left(\frac{1}{r}\right)^2\right) = 0,
   \label{eq:geo4b}
\end{equation}
this is the same function as \eqref{eq:geo4a} and certainly leads to the same result when solved, i.e.
\begin{equation}
    r(s) \approx m \pm s \sqrt{E^2-\epsilon }+c_1.
    \label{eq:um}
\end{equation}
As we said before, this shows that the function $r(s)$ is extensible to future infinity. For $r< < 1$ we have
\begin{equation}
   \frac{\pi  a^3 \left(-E^2\right)-\pi  a L^2+4 L^2 m}{\pi  a^3}+\epsilon  \left(1-\frac{4
   m}{\pi  a}\right)+\Dot{r}^2+\mathcal{O}\left(r^2\right) = 0,
   \label{eq:geo5b}
\end{equation}
note that \eqref{eq:geo5b} is slightly different from \eqref{eq:geo5a} in that it does not contain a term proportional to $1/r$ or the like. The integration leads us to
\begin{equation}
    r(s) \approx \pm \frac{s \sqrt{a^3 E^2+\frac{(4 m-\pi  a) \left(a^2 \epsilon -L^2\right)}{\pi }}}{a^{3/2}} + c_8.
    \label{eq:dois}
\end{equation}
In the above equation, $c_8$ is an integration constant. Now the function $r(s)$ admits any value of the affine parameter and is therefore extensible at the origin.
We will strengthen this result later using the Kretschmann scalar. For $r = r_+$ we have the same form of the geodesic equation as in the previous cases, and it can be reduced to
\begin{equation}
    A_3+A_4 \left(  r - r_+   \right) + \Dot{r}^2 + \mathcal{O}\left(\left(r-r_+\right)^2\right) = 0,
    \label{geo5}
\end{equation}
where $A_3$ and $A_4$ are
\begin{equation}
   A_3 = \frac{\pi r_+ \left(-\left(a^2+r_+^2\right) \left(E^2-\epsilon \right)-L^2\right)+4 m
   \tan ^{-1}\left(\frac{r_+}{a}\right) \left(L^2-\epsilon  \left(a^2+r_+^2\right)\right)}{\pi 
   r_+ \left(a^2+r_+^2\right)},
\end{equation}
\begin{equation}
    A_4 = \frac{4 m \tan ^{-1}\left(\frac{r_+}{a}\right) \left(\epsilon  \left(a^2+r_+^2\right)^2-L^2
   \left(a^2+3 r_+^2\right)\right)-4 a m r_+ \epsilon  \left(a^2+r_+^2\right)+2 L^2
   r_+ \left(2 a m+\pi  r_+^2\right)}{\pi  r_+^2 \left(a^2+r_+^2\right)^2}.
\end{equation}
Solving equation \eqref{geo5}, here  $c_9$ is the integration constant, lead us to
\begin{equation}
    r(s) \approx \biggl[ - \frac{ A_3}{A_4}   +r_+ - \frac{A_3}{4} \left( s \pm c_9  \right) \biggr].
    \label{eq:tres}
\end{equation}
This function $r(s)$ admits any value of $s$ and is therefore extensible over the event horizon. Summarizing the function $r(s)$ from these three limits \eqref{eq:um}, \eqref{eq:dois}, and \eqref{eq:tres}, we conclude that spacetime is geodesically complete.

Let us now show that the new Kretschmann scalar is finite everywhere. For \eqref{eq:ds2} it is
\begin{equation}
    \begin{aligned}
    K & = \frac{\frac{32 m^2 \left(\left(a^2+r^2\right) \tan ^{-1}\left(\frac{r}{a}\right)-a r\right)^2}{\pi ^2 r^2
   \left(a^2+r^2\right)^2}+\frac{64 m^2 \left(a^2+r^2\right)^2 \left(\frac{a r \left(a^2+2
   r^2\right)}{\left(a^2+r^2\right)^2}-\tan ^{-1}\left(\frac{r}{a}\right)\right)^2}{\pi ^2 r^6}+\frac{4
   \left(\pi  a^2+4 m r \tan ^{-1}\left(\frac{r}{a}\right)\right)^2}{\left(\pi  a^2+\pi 
   r^2\right)^2}}{\left(a^2+r^2\right)^2} \\ 
   &+\frac{2 \left(4 m \left(r^2-a^2\right) \tan
   ^{-1}\left(\frac{r}{a}\right)+2 a r (\pi  a-2 m)\right)^2}{\pi ^2 r^2 \left(a^2+r^2\right)^4}
    \end{aligned}
    \label{eq:quatro}
\end{equation}

This time there is no evidence of singularities in the geodesic analysis. However, we will check the asymptotic behavior at the origin, since this is the region where the scalar diverged in the previous section. If we expand \eqref{eq:quatro} around zero for $r < < 1$, we have
\begin{equation}
    K (r \sim 0) = \frac{4 \left(27 \pi ^2 a^2-144 \pi  a m+304 m^2\right)}{9 \pi ^2 a^6}-\frac{16 r^2 \left(45 \pi ^2 a^2-310
   \pi  a m+688 m^2\right)}{15 \left(\pi ^2 a^8\right)}+\mathcal{O}\left(r^3\right),
\end{equation}
with this result we can see that the Kretschmann scalar is finite at the origin and thus spacetime is regular everywhere. As we have already commented, this model appears in \cite{Lobo:2020ffi} and is called a black-bounce solution in this particular spacetime. We can reinterpret the figures \ref{fig:um} and \ref{fig:dois} (changing the function $\Sigma (r)$ does not change these graphs) in the context of black bounces.  
Information on how to obtain it and the details of its properties can be found in the aforementioned reference.

\subsection{Model 2}
\begin{figure}[htpb]
    \centering
    \includegraphics[scale=0.5]{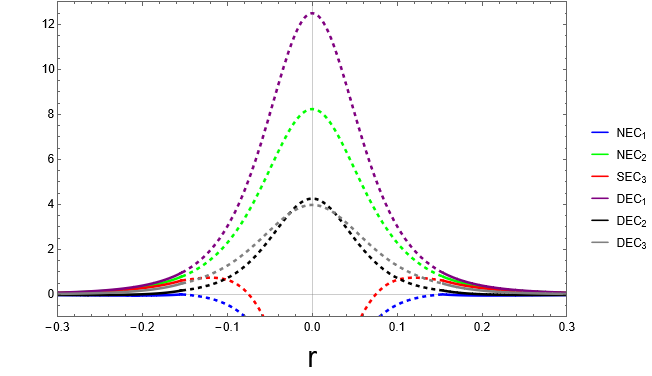}
    \caption{Graphical representation of the energy conditions for the regularized model 2 with $m=1$ and $a=0.01.$ The filled curves represent regions outside and the dashed curves represent regions within the event horizon.}
    \label{fig:energy5}
\end{figure}

The the spherically symmetric metric considered is 
\begin{equation}
    ds^2 = \Biggl[1-2 m \tan ^{-1}\left(\frac{a}{\sqrt{a^2+r^2}}\right) \Biggr]    dt^2 - \frac{dr^2}{\Biggl[1-2 m \tan ^{-1}\left(\frac{a}{\sqrt{a^2+r^2}}\right) \Biggr]  } - \left(a^2+r^2 \right)  \left( d \theta^2 + \sin^2 \theta d \phi^2   \right),
    \label{eq:ds3}
\end{equation}

\begin{equation}
    V(\Phi) = \frac{a m \, _2F_1\left(-\frac{3}{2},1;-\frac{1}{2};-\sec ^2\left(2 \sqrt{2 \pi } \Phi \right)\right)}{6 \pi 
   \left(a^2 \sec ^2\left(2 \sqrt{2 \pi } \Phi \right)\right)^{3/2}},
\end{equation}
and

\begin{equation}
    L(F) =\frac{m \left(F^2+F+\frac{3 (F+1) \left(F \tan ^{-1}\left(\sqrt{F}\right)+2 \cot
   ^{-1}\left(\sqrt{F}\right)\right)}{\sqrt{F}}+6\right)}{12 \pi  a (F+1) \sqrt{\frac{a^2}{F}}} .
\end{equation}
The function $_2F_1(a,b;c;z)$ is the  hypergeometric function defined by the series
\begin{equation}
   _2F_1(a,b;c;z) = \sum_{n=0}^{\infty} \frac{(a)_n (b)_n}{(c)_n} \frac{z^n}{n!},
\end{equation}
where $(a)_n$, $(c)_n$, and $(b)_n$ are the Pochhammer symbols.
We present the energy conditions for this case in the figure \ref{fig:energy5}. Note that again $NEC_1$ and $SEC_3$ can be violated within the event horizon.

With  \eqref{eq:ds3}  we get the following geodesic equation
\begin{equation}
  \Dot{r}^2+  \frac{2 m \tan ^{-1}\left(\frac{a}{\sqrt{a^2+r^2}}\right) \left(L^2-\epsilon 
   \left(a^2+r^2\right)\right)}{a^2+r^2}-\frac{L^2}{a^2+r^2}-E^2+\epsilon = 0.
\end{equation}

The Taylor series expansion for $r> > 1$ leads to the same function as \eqref{eq:geo4b}, therefore $r(s)$ is extensible into the infinite future.
For $r< < 1$ we have
\begin{equation}
\tan ^{-1}\left(1\right) \left(\frac{2 L^2 m}{a^2}-2 m \epsilon
   \right)-\frac{L^2}{a^2}-E^2+\Dot{r}^2+\epsilon +\mathcal{O}\left(r^2\right),
\end{equation}
which the integration leads to 
\begin{equation}
  r(s) \approx   \frac{2 a c_{10} \pm  s \sqrt{2 a^2 \left(2 E^2+\pi  m \epsilon -2 \epsilon \right)-2 L^2 (\pi  m-2)}}{2
   a}.
\end{equation}
Here $c_{10}$ is a constant. From this we conclude that the function is extensible at the origin.
For $r=r_+$ we have the same format as for the previous models

\begin{equation}
    A_5+A_6 \left(  r - r_+   \right) + \Dot{r}^2 + \mathcal{O}\left(\left(r-r_+\right)^2\right) = 0,
    \label{geo6}
\end{equation}
where
\begin{equation}
  A_5 =   \frac{2 m \tan ^{-1}\left(\frac{a}{\sqrt{a^2+r_+^2}}\right) \left(L^2-\epsilon 
   \left(a^2+r_+^2\right)\right)}{a^2+r_+^2}-\frac{L^2}{a^2+r_+^2}-E^2+\epsilon,
\end{equation}
\begin{equation}
\begin{aligned}
    A_6 = &\Biggl\{2 r_+ \Biggl[ a^5 m \epsilon -a^3 m \left(L^2-2 r_+^2 \epsilon \right)-2 L^2 m
   \sqrt{a^2+r_+^2} \left(2 a^2+r_+^2\right) \tan
   ^{-1}\left(\frac{a}{\sqrt{a^2+r_+^2}}\right)+2 a^2 L^2 \sqrt{a^2+r_+^2} \\
  & +L^2 r_+^2  \sqrt{a^2+r_+^2}+a m r_+^2 \left(r_+^2 \epsilon
   -L^2\right)\Biggr] \Biggr\} \times \Biggl[ \left(a^2+r_+^2\right)^{5/2} \left(2 a^2+r_+^2\right)\Biggr]^{-1}.
     \end{aligned}
\end{equation}
These results show that the geodesic equations are complete in the whole space-time.

Let us now finally compute the Kretschmann scalar. For \eqref{eq:ds3} we obtain

\begin{equation}
    \begin{aligned}
K &= \frac{8 a^2 \left(2 a^3+m r^2 \sqrt{a^2+r^2}-2 a m \left(2 a^2+r^2\right) \tan
   ^{-1}\left(\frac{a}{\sqrt{a^2+r^2}}\right)+a r^2\right)^2}{\left(a^2+r^2\right)^4 \left(2
   a^2+r^2\right)^2} \\ & +\frac{\frac{8 a^2 m^2 r^4}{\left(a^2+r^2\right)^3
   \left(\frac{a^2}{a^2+r^2}+1\right)^2}+\frac{4 \left(2 m r^2 \tan
   ^{-1}\left(\frac{a}{\sqrt{a^2+r^2}}\right)+a^2\right)^2}{\left(a^2+r^2\right)^2}+\frac{4 m^2 \left(2
   a^5-a^3 r^2-2 a r^4\right)^2}{\left(a^2+r^2\right) \left(2 a^2+r^2\right)^4}}{\left(a^2+r^2\right)^2}
    \end{aligned}
      \label{Kre2}
\end{equation}
which is finite at all points of space-time. To make it clearer, we expand \eqref{Kre2} in Taylor series for $r< < 1$ around the zero point
\begin{equation}
    K ( r \sim 0 ) = \frac{m \left(-8 \pi   +m+2 \pi ^2 m\right)+12}{a^4}+\frac{r^2 \left(4 (4+9 \pi )
   a m-2 a \left((3+4 \pi  (1+\pi )) m^2+24\right)\right)}{a^7}+\mathcal{O}\left(r^3\right).
\end{equation}
As we can see, there are no terms $1/r$ or the like. Once again we have achieved our original goal, we have regularized spacetime which previously had curvature singularities.

\subsection{Model 3}
For our last model the metric is given by
\begin{equation}
    ds^2 = \Biggl[1-\frac{2 m r}{a^2+r^2} \Biggr]    dt^2 - \frac{dr^2}{\Biggl[1-\frac{2 m r}{a^2+r^2} \Biggr]  } - \left(a^2+r^2 \right)  \left( d \theta^2 + \sin^2 \theta d \phi^2   \right),
    \label{eq:ds4}
\end{equation}

The potential associated with the scalar field $\Phi$   and the Lagrangian are

\begin{equation}
\begin{aligned}
   & V(\Phi) = \frac{m \left(21 \sin \left(4 \sqrt{2 \pi } \Phi \right)+6 \sin \left(8 \sqrt{2 \pi } \Phi \right)+\sin
   \left(12 \sqrt{2 \pi } \Phi \right)+24 \tan ^{-1}\left(\tan \left(2 \sqrt{2 \pi } \Phi
   \right)\right)\right)}{192 \pi  a^3},
   \end{aligned}
\end{equation}
and

\begin{equation}
    L(F) = \frac{m \left(\sqrt{F} (2 F+1) (4 F-3) \sqrt{-a^2 (F-1)}-3 a \tan ^{-1}\left(\frac{\sqrt{-a^2 (F-1)}}{a
   \sqrt{F}}\right)\right)}{24 \pi  a^4}.
\end{equation}

The energy conditions for the region outside the horizon are
\begin{equation}
    NEC_1 \rightarrow -\frac{a^2 \left(a^2+r (r-2 m)\right)}{4 \pi  \left(a^2+r^2\right)^3} \geq 0, \ \ \ \ \ NEC_2 \rightarrow  \frac{a^2 m r}{\pi  \left(a^2+r^2\right)^3} \geq  0
\end{equation}
\begin{equation}
    DEC_3 \rightarrow -\frac{a^2 \left(a^2+r (r-6 m)\right)}{8 \pi  \left(a^2+r^2\right)^3}  \geq 0  \ \ \ \ \ SEC_3 \rightarrow  \frac{a^2 m r}{\pi  \left(a^2+r^2\right)^3} \geq 0,
\end{equation}
\begin{equation}
    DEC_1 \rightarrow  \frac{a^2 m r}{\pi  \left(a^2+r^2\right)^3} \geq 0   \ \ \ \ \ DEC_2 \rightarrow   -\frac{a^2 \left(a^2+r (r-2 m)\right)}{4 \pi  \left(a^2+r^2\right)^3} \geq 0.
    \end{equation}
We see that $NEC_1$ and $DEC_{2,3}$ can be violated and $NEC_2$, $SEC_3$ and $DEC_1$ are are always satisfied.
For the region outside the horizon are
\begin{equation}
    NEC_1 \rightarrow \frac{a^2 \left(a^2+r (r-2 m)\right)}{4 \pi  \left(a^2+r^2\right)^3} \geq 0, \ \ \ \ \ NEC_2 \rightarrow  \frac{a^2 m r}{\pi  \left(a^2+r^2\right)^3} \geq  0
\end{equation}
\begin{equation}
    DEC_3 \rightarrow \frac{a^4+a^2 r (2 m+r)}{8 \pi  \left(a^2+r^2\right)^3}   \geq 0  \ \ \ \ \ SEC_3 \rightarrow \frac{a^2}{2 \pi  \left(a^2+r^2\right)^2} \geq 0,
\end{equation}
\begin{equation}
    DEC_1 \rightarrow  \frac{a^2 m r}{\pi  \left(a^2+r^2\right)^3} \geq 0   \ \ \ \ \ DEC_2  \rightarrow  0 \geq 0.
    \end{equation}
Now, the only condition that can be violated is $NEC_1$.

If we identify the components given by \eqref{eq:ds4} and then substitute them into  \eqref{eq:geo3}, we get
\begin{equation}
\Dot{r}^2+\frac{\left(\frac{2 m r}{a^2+r^2}-1\right) \left(E^2 \left(a^2+r^2\right)+L^2 \left(1-\frac{2 m
   r}{a^2+r^2}\right)\right)}{\left(a^2+r^2\right) \left(1-\frac{2 m r}{a^2+r^2}\right)}-\epsilon 
   \left(\frac{2 m r}{a^2+r^2}-1\right) = 0.
\end{equation}

Again, we will work with Taylor series expansions of this general geodesic equation. To $r> > 1$:
\begin{equation}
\Dot{r}^2 -E^2+\epsilon -\frac{2 m \epsilon }{r}+\mathcal{O}\left(\left(\frac{1}{r}\right)^2\right),
\end{equation}
a result which has already been obtained several times and, as already mentioned, allows the conclusion that the function $r(s)$ is extensible into the infinite future.
To $r< < 1$:
\begin{equation}
  \Dot{r}^2  -\frac{a^2 E^2+L^2}{a^2}+\epsilon +r \left(\frac{2 L^2 m}{a^4}-\frac{2 m \epsilon
   }{a^2}\right)+\mathcal{O}\left(r^2\right) = 0,
\end{equation}
the integration of the above function leads to two different results
\begin{equation}
    r(s) \approx \frac{c_{11} m s \left(L^2-a^2 \epsilon \right)}{a^2}-\frac{1}{2} c_{11}{}^2 m \left(\pm L^2 \mp a^2 \epsilon
   \right)+\frac{a^6+a^2 m^2 s^2 \epsilon +\frac{a^8 E^2}{L^2-a^2 \epsilon }-L^2 m^2 s^2}{2 a^4 m},
\end{equation}
where $c_{11}$ is the integration constant for both.  These functions are  extensible on the limit $r \rightarrow 0$.
To $r = r_+$ we have, once again, an expansion of the type
\begin{equation}
    A_7+A_8 \left(  r - r_+   \right) + \Dot{r}^2 + \mathcal{O}\left(\left(r-r_+\right)^2\right) = 0,
    \label{aaa}
\end{equation}
where $A_7$ and $A_8$ are
\begin{equation}
    A_7 = \frac{a^4 \left(\epsilon -E^2\right)-a^2 \left(2 r_+ \left(E^2 r_++m \epsilon
   -r_+ \epsilon \right)+L^2\right)+r_+^3 \left(E^2 (-r_+)-2 m \epsilon
   +r_+ \epsilon \right)+L^2 r_+ (2 m-r_+)}{\left(a^2+r_+^2\right)^2},
\end{equation}
\begin{equation}
    A_8 = \frac{2 m \epsilon  \left(r_+^4-a^4\right)+2 L^2 \left(a^2 (m+r_+)+r_+^2 (r_+-3
   m)\right)}{\left(a^2+r_+^2\right)^3}.
\end{equation}
Integrating \eqref{aaa} we get
\begin{equation}
    r(s) \approx \biggl[ - \frac{ A_7}{A_8}   +r_+ - \frac{A_7}{4} \left( s \pm c_{12}  \right) \biggr],
\end{equation}
where $c_{12}$ is another constant of integration. The above function admits any value of $s$ and is therefore extensible in this limit. From this we can conclude that spacetime is now geodesically complete at all points, including the origin.

To complete the analysis, we compute the Kretschmann scalar to check for singularities. It is
\begin{equation}
   K = \frac{\frac{8 m^2 r^2 \left(r^2-a^2\right)^2}{\left(a^2+r^2\right)^4}+\frac{16 m^2 \left(r^3-3 a^2
   r\right)^2}{\left(a^2+r^2\right)^4}+4 \left(1-\frac{r^2 \left(1-\frac{2 m
   r}{a^2+r^2}\right)}{a^2+r^2}\right)^2+\frac{8 \left(a^4+a^2 r (r-3 m)+m
   r^3\right)^2}{\left(a^2+r^2\right)^4}}{\left(a^2+r^2\right)^2},
\end{equation}
which is now finite at all points. The asymptotic behavior near the origin is shown by the expansion below
\begin{equation}
  K (r \sim 0 ) = \frac{12}{a^4}-\frac{48 m r}{a^6}+\frac{r^2 \left(224 m^2-48 a^2\right)}{a^8}+\mathcal{O}\left(r^3\right).
\end{equation}
Therefore, we have regularized the spacetime of all three metric functions presented in section \ref{sec3}.


\section{Conclusion
\label{sec5}}
In this paper we point out that a regular metric is no guarantee that spacetime is also regular.
We believe that the way regularity is treated in some papers can confuse an inexperienced reader, and that is the motivation for this note.
We are aware that there are several definitions of regularity/singularity, so we have focused on a static and spherically symmetric spacetime that may or may not be asymptotically
flat. We emphasize here that singularity curvature is verified when one or more components of the Riemann tensor are infinite. The conditions given by \eqref{eq:con1} and \eqref{eq:con2} are not sufficient to guarantee this. As shown by\cite{Lobo:2020ffi}, the proof of the fitness of the Kretschmann scalar is sufficient to guarantee the regularity of spacetime. We have given three examples to illustrate our point. We study for each of them: the regularity condition of the metric, the event horizon and the extreme solution, the geodesic equation, and the Kretschmann scalar.
We show in the appendix \ref{sec:source} how to obtain a matter source for a static, spherically symmetric metric.
 Thus, we show that the models used here are exact solutions of Einstein's equations.

In the section \ref{sec:geo} we present a summary of the analysis of trajectories of free particles, i.e. geodesics. We give for the considered spacetime (static and spherically symmetric) the equation of a light particle moving in the equatorial plane. In section \ref{sec3} we have been engaged in showing examples of metrics satisfying the conditions \eqref{eq:con1} and \eqref{eq:con2}. But they produce a spacetime with curvature singularities.
We analyze for each case: the properties of the metric, the source of matter, the energy conditions, the geodesic equations, and the Kretschmann scalar.
The model in \ref{subsec1} was first described in \cite{Lobo:2020ffi} as a black bounce. It has two symmetric event horizons that become one when $a > 4m/\pi$.Model two, presented in section \ref{subsec2}, also has two symmetric event horizons, but for this metric there is no extremization condition for parameter $a$. The third model, seen in
\ref{subsec3}, has two positive event horizons and an extremization condition for the case where $a > m$. We use the method described in the appendix to find a source for the solutions. We show that they can be obtained from general relativity minimally coupled to nonlinear electrodynamics. The equations \eqref{L1}, \eqref{L2}, and \eqref{L3} are the Lagrangians of models 1, 2, and 3, respectively.
We note that the geodesics of all examples are inextensible at the origin, a behavior associated with the presence of a curvature singularity. We emphasize this using the Kretschmann scalar and show  that the spacetime generated by these metrics is indeed singular.

In \ref{sec4} we use the Simpson-Visser model of the black-bounce model, $ \Sigma(r) = \sqrt{a^2+r^2}$,for the regularization of the space-time of
 the metrics under consideration. We have found that in this spacetime, nonlinear electrodynamics alone is not sufficient to serve as a source. It is necessary to add a scalar field $\Phi (r)$ associated with a potential $V(r)$. The scalar field depends only on $\Sigma (r)$, so the equation \eqref{Phi} is valid for the three metrics. We calculate the potential $V(r)$ for each model and have also recalculated the Lagrangian and energy conditions.
 For verification, we recalculated the geodesic equations and showed that they are now complete for every point in space. Furthermore, the Kretschmann scalar became finite after this change.
 
Finally, we would like to point out that models 2 and 3 are new metrics. Given the properties of the event horizons, these models can potentially be studied as black-bounce solutions. We intend to study the physical properties and causal structure of these new solutions.
We also plan to study the thermodynamics, scalar wave absorption and scattering, and shadows of singular and regular solutions in a future work.

 
\vspace{1cm}

{\bf Acknowledgements}: M. E. R.  thanks Conselho Nacional de Desenvolvimento Cient\'ifico e Tecnol\'ogico - CNPq, Brazil, for partial financial support. This study was financed in part by the Coordena\c{c}\~{a}o de Aperfei\c{c}oamento de Pessoal de N\'{i}vel Superior - Brasil (CAPES) - Finance Code 001.

\newpage
\appendix
\section{Sources
\label{sec:source}}
In this appendix we will show that the metrics used in this paper are solutions of the Einstein equations. To this end, we propose a source for each model.
First, we consider a general metric for a static and spherically symmetric spacetime, which is as follows
\begin{equation}
    ds^2 = f(r) dt^2 - \frac{dr^2}{f(r)} - \Sigma^2(r) \left( d \theta^2 + \sin^2 \theta d \phi^2   \right),
\end{equation}
where $f(r)$ and $\Sigma(r)$ are arbitrary functions of the variable r. Then we assume that the theory is described by the action \cite{Rodrigues:2023vtm}
\begin{equation}
    S = \int d^4x \sqrt{-g} \left[ R - 2 \kappa \left( \epsilon g^{\mu \nu } \partial_{\mu} \Phi \partial_{\nu} \Phi - V(\Phi)  \right) + \kappa \mathcal{L}(F) \right].
    \label{eq:acaoA}
\end{equation}
In the above equation
 $R$ is the Ricci scalar, $g_{\mu \nu}$ are the components of the metric tensor, $g$ is the determinant of the metric, $\kappa =8 \pi$ (in the geometrized system of units), $\Phi$ is a scalar field, $V (\Phi)$ is the potential associated with the scalar field, $\epsilon = \pm 1$ is the type of the field ($+1$ is the usual scalar field and $-1$ is a phantom field), and $L(F)$ is the nonlinear Lagrangian function of the scalar $F = F^{\mu \nu}F_{\mu \nu}/4$.
 
 From the proposed action \eqref{eq:acaoA} we have the following field equations
 \begin{equation}
     \nabla_{\mu} \left(  L_F F^{\mu \nu }\right) =0,
     \label{eq:camponed}
 \end{equation}
 \begin{equation}
      2 \epsilon \nabla_{\mu} \nabla^{\mu} \Phi = \frac{d V}{d \Phi},
     \label{eq:camposcalar}
 \end{equation}
\begin{equation}
    R_{\mu \nu} - \frac{1}{2} g_{\mu \nu}R  = \kappa T_{\mu \nu}^{NED}  + \kappa T_{\mu \nu}^{\Phi}.
    \label{eqeinsteinc}
\end{equation}
Here $T_{\mu \nu}^{ NED }$ is the stress-energy tensor of the matter sector, which is defined for nonlinear electrodynamics (NED) as follows
 \begin{equation}
T_{\mu \nu}^{NED} = g_{\mu \nu} L - L_F  F_{\mu}^{\ \alpha} F_{\nu \alpha},
\end{equation}
with $L_F = \partial L/ \partial F$, and $T_{\mu \nu}^{\Phi}$ is the stress-energy tensor of the scalar field sector given by 
\begin{equation}
    T_{\mu \nu}^{\Phi} = 2 \epsilon  \partial_{\mu} \Phi \partial_{\nu} \Phi - g_{\mu \nu}\left(  \epsilon g^{\alpha \beta } \partial_{\alpha} \Phi \partial_{\beta} \Phi  -V(\Phi) \right) .
\end{equation}
In addition, we will focus on magnetic solutions, hat, the only nonzero component of tensor $F_{\mu \nu}$ is
\begin{equation}
    F_{23} = q \sin \theta,
\end{equation}
where $q$ is a self-gravitating magnetic charge described by nonlinear electrodynamics.
Then the electromagnetic scalar
\begin{equation}
    F(r) = \frac{q^2}{2 \Sigma^4}.
\end{equation}
With that choices, the nontrivial field equations are
\begin{equation}
    \begin{aligned}
        & \frac{\Sigma  \left(f' \Sigma '+2 f \Sigma ''\right)+f \Sigma '^2-1}{\Sigma ^2} = -\kappa  \left(\epsilon  f \Phi '^2+L+V\right) \\
        & \frac{\Sigma  f' \Sigma '+f \Sigma '^2-1}{\Sigma ^2} = - \kappa  \left(V-\epsilon  f \Phi '^2+L\right) \\
        & \frac{\Sigma  f''+2 f' \Sigma '+2 f \Sigma ''}{2 \Sigma } = -\kappa  \left(L-\frac{q^2 L_F}{\Sigma ^4}+\epsilon  f \Phi '^2+V\right) \\
        &2 \epsilon  \left(f' \Phi '+f \Phi ''\right)+\frac{4 \epsilon  f \Sigma ' \Phi
   '}{\Sigma }=\frac{V'}{\Phi '}.
    \end{aligned}
    \label{eq:fieldeq}
\end{equation}
The prime denotes a derivative with respect to the $r$-coordinate. So the Lagrangian $L(F)$, its derivative with respect to the electromagnetic scalar $L_F$, the scalar field $\Phi (r)$ and the associated potential $V(\Phi)$ are respectively

\begin{equation}
    L = \frac{1-\Sigma ' \left(\Sigma  f'+f \Sigma '\right)}{\kappa  \Sigma ^2}+\epsilon  f
   \Phi '^2-V,
\end{equation}
\begin{equation}
    L_F = \frac{\Sigma ^2 \left(\Sigma ^2 f''+f \left(2 \Sigma  \left(\Sigma ''+2 \kappa  \epsilon 
   \Sigma  \Phi '^2\right)-2 \Sigma '^2\right)+2\right)}{2 \kappa  q^2},
\end{equation}
\begin{equation}
    \Phi^{\prime}  = \frac{i}{ \sqrt{\kappa \epsilon}} \sqrt{\frac{\Sigma^{\prime \prime}}{\Sigma}},
\end{equation}
\begin{equation}
    V^{\prime} = \frac{ 2 \Phi^{\prime} \epsilon \left(  \Sigma f^{\prime} \Phi^{\prime} + 2  f \Sigma^{\prime} \Phi^{\prime} +  f \Sigma \Phi^{\prime \prime} \right)}{\Sigma}.
\end{equation}

Using this procedure, we can specify $f(r)$ and $\Sigma (r)$ and obtain a suitable source for each solution. It is worth noting that we do not need to use a coupled scalar field in the \ref{sec3} section, this is only necessary \cite{Rodrigues:2023vtm} if we are dealing with Simpson-Visser spacetime (i.e. for the \ref{sec4} section).

\section{Invariants \label{sec:invariantes}}

We know that there are several other invariants in general relativity besides the Kretschmann scalar. The Ricci scalar $ R_{\mu \nu}g^{\mu \nu}$ and the Ricci squared $ R_{\mu \nu} R^{\mu \nu}$ are commonly used in the verification of curvature singularities. For a static and spherically symmetric spacetime described by \eqref{eq:ds}, they are as follows

\begin{equation}
\begin{aligned}
   & R_{\mu \nu} R^{\mu \nu} =\frac{1}{2 \Sigma (r)^4} \Bigl[ \Sigma (r)^4 f''(r)^2+4 \Sigma (r)^2 \left(2 f'(r)^2 \Sigma '(r)^2+4 f(r) f'(r) \Sigma '(r) \Sigma ''(r)+3
   f(r)^2 \Sigma ''(r)^2\right) \\ &+8 \Sigma (r) \left(f(r) \Sigma '(r)^2-1\right) \left(f'(r) \Sigma '(r)+f(r) \Sigma
   ''(r)\right) +4 \Sigma (r)^3 f''(r) \left(f'(r) \Sigma '(r)+f(r) \Sigma ''(r)\right)+4 \left(f(r) \Sigma
   '(r)^2-1\right)^2 \Bigr],
   \end{aligned}
   \label{eq:Ricci2}
\end{equation}
and
\begin{equation}
    R_{\mu \nu} g^{\mu \nu} = \frac{\Sigma (r)^2 f''(r)+4 \Sigma (r) \left(f'(r) \Sigma '(r)+f(r) \Sigma ''(r)\right)+2 f(r) \Sigma '(r)^2-2}{\Sigma
   (r)^2}.
   \label{eq:scalarR}
\end{equation}
But note that we do not use it here for the sake of practicality. Take for example the metric \eqref{eq:metrica1}, the invariants \eqref{eq:Ricci2} and \eqref{eq:scalarR} are

\begin{equation}
    R_{\mu \nu} R^{\mu \nu} =  \frac{32 a^2 m^2 \left(a^4+2 a^2 r^2+2 r^4\right)}{\pi ^2 r^4 \left(a^2+r^2\right)^4},
\end{equation}
\begin{equation}
    R_{\mu \nu} g^{\mu \nu} = -\frac{8 a^3 m}{\pi  r^2 \left(a^2+r^2\right)^2}.
\end{equation}
In this case, both diverge at $r \rightarrow 0$. This is not new information, we found this out while checking the asymptotic behavior of the Kretschmann scalar.

\section{Energy conditions \label{sec:energy}}
In this appendix, we  define the energy conditions that we  used to analyze the solutions. We follow the definitions used in \cite{Lobo:2020ffi}, which are more appropriate for the case we are dealing with. First, we adopt the line element
\begin{equation}
    ds^2 = f(r) dt^2 - \frac{dr^2}{f(r)} - \Sigma^2(r) \left( d \theta^2 + \sin^2 \theta d \phi^2   \right),
\end{equation}
where $f(r)$ and $\Sigma(r)$ are arbitrary functions of the variable $r$.
For this particular line element, the Einstein equations
\begin{equation}
    R_{\mu \nu} - \frac{1}{2}  g_{\mu \nu} R = \kappa^2 T_{\mu \nu},
\end{equation}
where we use the convention that
\begin{equation}
    T^{\mu}_{\ \nu} = \text{diag}[\rho, -p_1,-p_2,-p_2],
\end{equation}
lead to the following stress-energy profile (outside the event horizon, i.e. t is a timelike coordinate)
\begin{equation}
    \rho = -\frac{\Sigma (r) \left(f'(r) \Sigma '(r)+2 f(r) \Sigma
   '(r)^2\right)+f(r) \Sigma '(r)^2-1}{\kappa ^2 \Sigma (r)^2},
\end{equation}
\begin{equation}
    p_1 = \frac{\Sigma (r) f'(r) \Sigma '(r)+f(r) \Sigma '(r)^2-1}{\kappa ^2,
   \Sigma (r)^2}
\end{equation}
\begin{equation}
    p_2 = \frac{\Sigma (r) f''(r)+f'(r) \Sigma ''(r)+2 f'(r) \Sigma '(r)}{2
   \kappa ^2 \Sigma (r)}.
\end{equation}
For  inside the event horizon, t is spacelike coordinate and we have
\begin{equation}
     T^{\mu}_{\ \nu} = \text{diag}[-p1, \rho,-p_2,-p_2],
\end{equation}
\begin{equation}
    \rho = \frac{-\Sigma (r) f'(r) \Sigma '(r)-f(r) \Sigma '(r)^2+1}{\kappa ^2
   \Sigma (r)^2},
\end{equation}
\begin{equation}
    p_1 = \frac{\Sigma (r) \left(f'(r) \Sigma '(r)+2 f(r) \Sigma
   '(r)^2\right)+f(r) \Sigma '(r)^2-1}{\kappa ^2 \Sigma (r)^2},
\end{equation}
\begin{equation}
    p_2 = \frac{\Sigma (r) f''(r)+f'(r) \Sigma ''(r)+2 f'(r) \Sigma '(r)}{2
   \kappa ^2 \Sigma (r)}.
\end{equation}
The energy conditions are
\begin{equation}
      NEC_{1,2} = WEC_{1,2} = SEC_{1,2} \rightarrow \rho + p_{1,2} \geq 0,
\end{equation}
\begin{equation}
SEC_3 \rightarrow \rho + p_1 +2p_2 \geq 0,
\end{equation}
\begin{equation}
     DEC_{1,2} \rightarrow \rho  - |p_{1,2}| \geq 0,
\end{equation}
\begin{equation}
     DEC_3 = WEC_3  \rightarrow \rho \geq 0.
\end{equation}

In therms of the arbitrary functions $f(r)$ and $\Sigma(r)$ the energy conditions outside of the event horizon are
\begin{equation}
    NEC_1 = WEC_1 = SEC_1 \rightarrow -\frac{2 f(r) \Sigma ''(r)}{\kappa ^2 \Sigma (r)} \geq 0 ,
\end{equation}
\begin{equation}
    NEC_2 = WEC_2 = SEC_2  \rightarrow \frac{\Sigma (r)^2 f''(r)-2 f \left(\Sigma (r) \Sigma ''(r)+\Sigma
   '(r)^2\right)+2}{2 \kappa ^2 \Sigma (r)^2} \geq 0,
\end{equation}
\begin{equation}
SEC_3 \rightarrow   \frac{\Sigma (r) f''(r)+2 f'(r) \Sigma '(r)}{\kappa ^2 \Sigma (r)} \geq 0;
\end{equation}
\begin{equation}
    DEC_1 \rightarrow   \frac{2 \left(-\Sigma (r) f'(r) \Sigma '(r)-f(r) \Sigma (r) \Sigma
   ''(r)-f(r) \Sigma '(r)^2+1\right)}{\kappa ^2 \Sigma (r)^2} \geq 0 
\end{equation}
\begin{equation}
  DEC_2  \rightarrow    -\frac{\Sigma (r)^2 f''(r)+\Sigma (r) \left(4 f'(r) \Sigma '(r)+6
   f(r) \Sigma ''(r)\right)+2 f(r) \Sigma '(r)^2-2}{2 \kappa ^2
   \Sigma (r)^2} \geq 0 
\end{equation}
\begin{equation}
 DEC_3 = WEC_3  \rightarrow   -\frac{\Sigma (r) \left(f'(r) \Sigma '(r)+2 f(r) \Sigma ''(r)\right)+f(r) \Sigma
   '(r)^2-1}{\kappa ^2 \Sigma (r)^2} \geq 0.
\end{equation}

Inside of the event horizon they are
\begin{equation}
    NEC_1 = WEC_1 = SEC_1 \rightarrow \frac{2 f(r) \Sigma ''(r)}{\kappa ^2 \Sigma (r)} \geq 0,
\end{equation}
\begin{equation}
     NEC_2 = WEC_2 = SEC_2  \rightarrow  \frac{\Sigma (r)^2 f''(r)-2 f \left(\Sigma (r) \Sigma ''(r)+\Sigma
   '(r)^2\right)+2}{2 \kappa ^2 \Sigma (r)^2} \geq 0,
\end{equation}
\begin{equation}
    SEC_3 \rightarrow \frac{\Sigma (r) f''(r)+2 f'(r) \Sigma '(r)+4 f(r) \Sigma ''(r)}{\kappa ^2 \Sigma (r)} \geq 0,
\end{equation}
\begin{equation}
     DEC_1 \rightarrow  \frac{2 \left(-\Sigma (r) f'(r) \Sigma '(r)-f(r) \Sigma (r) \Sigma
   ''(r)-f(r) \Sigma '(r)^2+1\right)}{\kappa ^2 \Sigma (r)^2} \geq 0,
\end{equation}
\begin{equation}
      DEC_2  \rightarrow -\frac{\Sigma (r)^2 f''(r)+\Sigma (r) \left(4 f'(r) \Sigma '(r)+2
   f(r) \Sigma ''(r)\right)+2 f(r) \Sigma '(r)^2-2}{2 \kappa ^2
   \Sigma (r)^2}  \geq 0,
\end{equation}
\begin{equation}
     DEC_3 = WEC_3  -\frac{\Sigma (r) f'(r) \Sigma '(r)+f(r) \Sigma '(r)^2-1}{\kappa ^2 \Sigma (r)^2}   \geq 0.
\end{equation}
Using these definitions, we can see that some conditions are the same both inside and outside the horizon. However, in the special case of $\Sigma^{\prime \prime}(r) = 0$, there is no difference in the energy conditions between these two regions.


\end{document}